\documentclass[aps,prl,twocolumn,superscriptaddress]{revtex4}

\usepackage{amsmath,amssymb,amsfonts}
\usepackage{graphicx}

\usepackage{wrapfig}

\begin{document}

\title{Twists and turns for metamaterials}

\author{Mingkai Liu}
\affiliation{Nonlinear Physics Centre and Centre for Ultrahigh-bandwidth Devices for Optical Systems (CUDOS),  Research School of Physics and Engineering, Australian National University, Canberra ACT 0200, Australia}

\author{Yue Sun}
\affiliation{Nonlinear Physics Centre and Centre for Ultrahigh-bandwidth Devices for Optical Systems (CUDOS),  Research School of Physics and Engineering, Australian National University, Canberra ACT 0200, Australia}

\author{David A.~Powell}
\affiliation{Nonlinear Physics Centre and Centre for Ultrahigh-bandwidth Devices for Optical Systems (CUDOS),  Research School of Physics and Engineering, Australian National University, Canberra ACT 0200, Australia}

\author{Ilya V.~Shadrivov}
\affiliation{Nonlinear Physics Centre and Centre for Ultrahigh-bandwidth Devices for Optical Systems (CUDOS),  Research School of Physics and Engineering, Australian National University, Canberra ACT 0200, Australia}

\author{Mikhail Lapine}
% \email[Corresponding author: ]{mlapine@physics.usyd.edu.au}
\affiliation{Centre for Ultrahigh-bandwidth Devices for Optical Systems (CUDOS), School of Physics, University of Sydney, Sydney NSW 2006, Australia}

\author{Ross C.~McPhedran}
\affiliation{Centre for Ultrahigh-bandwidth Devices for Optical Systems (CUDOS), School of Physics, University of Sydney, Sydney NSW 2006, Australia}

\author{Yuri S.~Kivshar}
\affiliation{Nonlinear Physics Centre and Centre for Ultrahigh-bandwidth Devices for Optical Systems (CUDOS),  Research School of Physics and Engineering, Australian National University, Canberra ACT 0200, Australia}

\begin{abstract}
We propose and verify experimentally a new concept for achieving strong nonlinear coupling between the electromagnetic and elastic properties in metamaterials. This coupling is provided through a novel degree of freedom in metamaterial design: internal rotation within structural elements. Our meta-atoms have high sensitivity to electromagnetic wave power, and the elastic and electromagnetic properties can be independently designed to optimise the response. We demonstrate a rich range of nonlinear phenomena including self-tuning and bistability, and provide a comprehensive experimental demonstration of the predicted effects.
\end{abstract}

%\pacs{}

\maketitle

%-------------------------------------------------------------
% \section*{Introduction}
%-------------------------------------------------------------

Metamaterials research has grown rapidly over the past decade, exhibiting a wide variety of
new wave phenomena \cite{SolSha,MarMarSor}.
Being initially conceived in the domain of electromagnetics \cite{SPV0,PenPW,Sha07},
the metamaterial concept also proved to be fruitful in other areas of physics 
\cite{LiCha04,Nor08,NicMot12}.
Until recently, however, direct interplay between \emph{different types of physical effects within the same
metamaterial} was not considered, although mechanical
control over electromagnetic metamaterial properties was employed in structural tuning
\cite{LapPowGor09,LiuZhuTsa12}.

It turns out that introducing a mechanical degree of freedom into
electromagnetic metamaterials leads to an interesting range of nonlinear effects,
giving rise to a new class of magnetoelastic metamaterials \cite{LapShaPow12}
and to wide-band operation \cite{LapShaKiv12}.
The range of possible effects achievable in this way promises to be richer than in the prominent
area of optomechanics \cite{MarGir09}, because the greater flexibility in metamaterial
design overcomes the limits of available material functionalities, and offers wider possibilities
for optimisation.
At the same time, the implementation of magnetoelastic metamaterials \cite{LapShaPow12}
remains challenging and in some cases, such as the conformational nonlinearity in
resonant spirals \cite{LapShaPow11}, remains inaccessible for optics.
The reason for this is that the magnetic forces, employed in the initial designs,
are relatively weak, so such materials require either high power or extremely small elastic restoring forces, which poses a considerable manufacturing challenge.

We recall, however, that earlier research on structurally tunable metamaterials
\cite{LapPowGor09} indicated that near-field interaction may significantly improve the
tunability range, leading to various effects associated with
near-field coupling \cite{PowLapGor10}.
In particular, changing the mutual orientation between the neighbouring elements
has a profound effect on the resonator coupling and the structure of their modes \cite{PowHanSha11,LiuLiuLi09}.

We therefore expect that the most efficient approach to implement dynamic coupling
between electromagnetic and mechanical effects should rely on near-field interaction, which can have a powerful influence even for subtle changes in
the mutual orientation. Instead of a considerable displacement of an
entire array \cite{LapPowGor09}, it is
sufficient to move the crucial \emph{parts} of the resonant particles with respect
to each other --- for example, the gaps of the two coupled split-ring resonators ---
which involves a more gentle geometric alteration.

\begin{figure}[b]
\includegraphics[width=0.99\columnwidth]{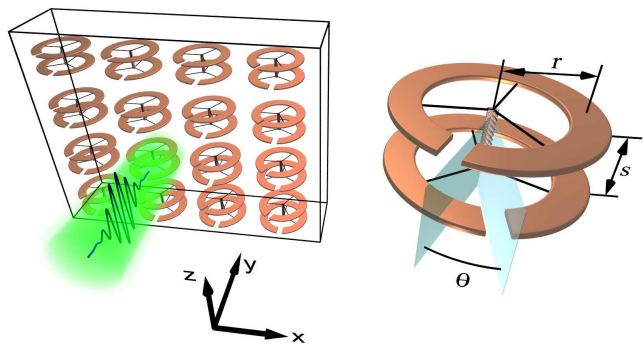}
\caption{Conceptual layout of a new metamaterial and its rotational ``meta-atom''. 
Incident wave propagates along $y$ direction, having a linear polarisation with
the electric field along $x$ and magnetic field along $z$.
The induced electromagnetic torque between the resonators changes
the mutual twist angle $\theta$ between the rings,
connected by an elastic wire.
\label{fig:scheme} }
\end{figure}

% 
% \section*{Results}
%
%-------------------------------------------------------------
% \subsection*{Design and general concept}
%-------------------------------------------------------------

Let us now introduce our novel concept: {\em nonlinear metamaterials with intrinsic rotation}.
As a building element of the structure (see Fig.~\ref{fig:scheme}), we consider two coaxial split ring resonators (SRRs) with elastic feedback between them. The rings are allowed to rotate about the common axis, while the
elastic feedback is provided by connecting a thin elastic wire. Indeed, the use of elastic wires has a prominent history in physics, being utilised in
the milestone achievement of the experimental demonstration of light pressure
by P.\,N.~Lebedev \cite{Leb01}. 
Here, we employ the electromagnetic (EM) torque to construct a ``light-driven'' meta-atom, which enables us to modulate the resonant frequencies by twisting the rotatable element \emph{directly} with EM waves. The component of the EM forces which twists the rings with respect
to each other is normally not the strongest among the forces involved, but the prominent advantage of using EM torque instead of collinear EM force to drive a meta-atom is that the effective lever arm of the azimuthal EM force can be much larger than that of the azimuthal restoring force from a thin wire; this can effectively magnify the deformation in azimuthal direction by orders of magnitude. 

Technically, there are a number of ways to implement this general scheme; in our design,
one of the rings is fixed to a substrate and the other one is suspended on the long wire.
The three symmetrically positioned wires which attach the suspended ring to the
string  provide stability against tilt.
In this design, therefore, the only favourable movement is the rotation of the suspended
ring with respect to the fixed ring over the common axis, and all other mechanical
degrees of freedom can be neglected.

Suppose the initial position is such that the ring slits have a certain angle between them with respect to the common axis (Fig.~\ref{fig:scheme}).
An EM wave then induces a certain distribution of charges and currents in the two
resonators, and the resonance of the system is determined by their mutual orientation \cite{PowHanSha11}.
These charges and currents also result in EM torque between the two rings \cite{LiuPowSha12}, which drives the suspended ring to rotate until the
EM torque is compensated by the elasticity of the twisted wire.
Meanwhile, the entire pattern
of charges and currents gets modified and the torque also changes, so the final stable equilibrium
is only achieved via a complex nonlinear feedback. The twisted wire provides a restoring torque to balance the EM torque, so that we can control the twist angle (and thus the resonance) by changing the external field signal.

An additional feature of the proposed design is that the nonlinear dynamics of the rotatable
particle not only depend on the parameters of the wire, but also on the EM mode initially
excited in the resonators, which is determined by the starting angle between the gaps.
The latter can be made arbitrary, and an implementation has the possibility
to deliberately adjust it, introducing tunability to the system.

The above design, therefore, offers \emph{a tunable resonant nonlinear system with elastic
feedback} and, as we show below, yields a rich pattern of nonlinear response including self-tuning and nonlinear bistability, which are much stronger than those provided by using nonlinear semiconductor components. Below, we present a detailed theoretical analysis of rotational meta-atoms. We then proceed to the experimental results obtained with a fabricated prototype of the rotational ``meta-atom'' placed in a rectangular waveguide, and confirm all important features predicted by theory. Finally, we perform full-wave numerical simulations of the array of ``meta-atoms'' (i.e., metamaterial), and demonstrate that all the nonlinear effects observed for the single meta-atom in the waveguide are qualitatively the same in the array.

%-------------------------------------------------------------
% \subsection*{Theoretical analysis}
%-------------------------------------------------------------

To start with, we use a semi-analytical model to study the dynamics of an isolated meta-atom in free space. It will subsequently be demonstrated that this model explains all qualitative features of an array.
As shown in Fig.~\ref{fig:scheme}, the two coaxial identical SRRs are offset by a distance $s$ in the z direction, and the twist angle between them is $\theta$. The incident wave propagates along the y direction, with its magnetic field in the z direction and electric field in the x direction. To study the nonlinear behaviour of the rotatable meta-atom, we utilise an efficient analytical model based on the single mode approximation and the near-field interaction \cite{PowHanSha11}. This model can provide a reasonable prediction of the EM response as well as the optomechanical properties of the structures \cite{LiuPowSha12}. As our previous studies showed \cite{LiuPowShaKiv12}, the current and charge of the SRR can be separated into frequency-dependent mode amplitudes and spatially-dependent distributions:
$\mathbf{J}(\mathbf{r},\omega)= -j\omega Q(\omega)\mathbf{j}(\mathbf{r})$ and $\rho(\mathbf{r},\omega)=Q(\omega)q(\mathbf{r})$.
The mode amplitudes $Q_{1,2}$ can be obtained after solving the coupled equations:
\begin{eqnarray}\label{eq:Q1}
\begin{split}
Q_{1}=(\mathcal{E}_{2}F_\mathrm{m}-\mathcal{E}_{1}F_\mathrm{s})/(F_\mathrm{s}^{2}-F_\mathrm{m}^{2})\\
Q_{2}=(\mathcal{E}_{1}F_\mathrm{m}-\mathcal{E}_{2}F_\mathrm{s})/(F_\mathrm{s}^{2}-F_\mathrm{m}^{2}),%\label{eq:Q2}
\end{split}
\end{eqnarray}
where $\mathcal{E}_{1}$ and $\mathcal{E}_{2}$ correspond to the effective voltage applied to the lower (index 1) and top (index 2) SRRs by the external fields; $F_\mathrm{s}$ and $F_\mathrm{m}$ are the self and mutual impedance terms, with their explicit forms given in the supplemental materials \cite{supp}.

Once the frequency-dependent mode amplitudes are known, we can calculate the EM torque experienced by the SRRs. Here, we are particularly interested in the torque on the top rotating ring:
$\mathbf{M}_\mathrm{EM}=\int_{V_{2}}\rho(\mathbf{r}_{2})\mathbf{r}_{2}\times \mathbf{E}+\mathbf{r}_{2}\times[\mathbf{J}(\mathbf{r}_{2})\times \mathbf{B}]\mathrm{d}V_{2}$, where the integration is performed over the volume $V_{2}$ of the top SRR.
We decompose the total torque into two parts: external torque $\mathbf{M}_\mathrm{ext}$ contributed by the external incident fields \cite{LiuPowSha12}, and internal torque $\mathbf{M}_\mathrm{int}$ due to the near-field interaction between the two SRRs. We assume that the lower SRR is fixed while the top SRR is only allowed to rotate about the z axis. See the supplemental materials \cite{supp} for explicit expressions for the internal and external torque.

\begin{figure}[b]
\includegraphics[width=0.99\columnwidth]{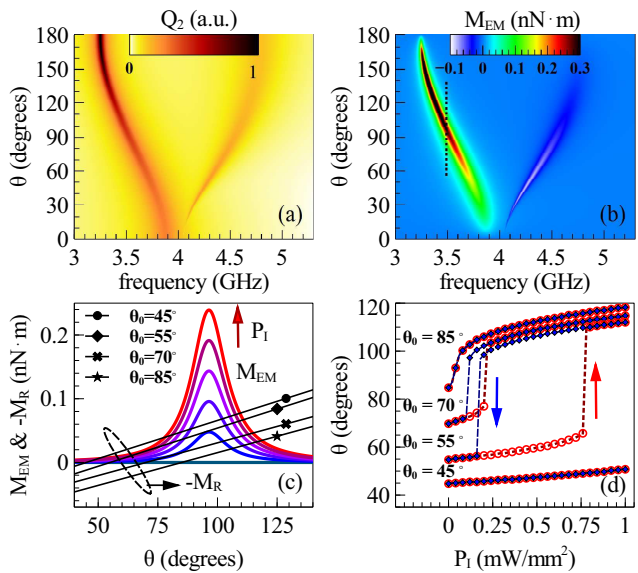}
\caption{ The principle of nonlinear response in rotatable meta-atoms.
(a) The mode amplitude $Q_{2}$ and (b) the EM torque $\mathbf{M}_\mathrm{EM}$ of the top rotatable ring. (c) The EM torque at 3.5\,GHz for different pump powers from 0 to 1\, mW/mm$^2$ in 0.2\,mW/mm$^2$ steps, and the restoring torque for different initial twist angle $\theta_{0}$; (d) the corresponding paths of power-dependent twist angles under different $\theta_{0}$.
\label{fig:bist} }
\end{figure}

Figures~\ref{fig:bist}(a) and ~\ref{fig:bist}(b) depict the mode amplitude $Q_{2}$ and the total EM torque $\mathbf{M}_\mathrm{EM}=\mathbf{M}_\mathrm{ext}+\mathbf{M}_\mathrm{int}$ experienced by the top SRR as functions of frequency and twist angle $\theta$. Since radiation losses are taken into account in the model \cite{LiuPowShaKiv12}, we are able to accurately describe the evolution of the mode amplitudes, phases and lineshapes of the resonances. As expected, this chiral meta-atom supports two hybrid resonances, which can be characterised as symmetric (lower frequency branch) and antisymmetric (higher frequency branch) modes, according to the symmetry of the $H_{z}$ component \cite{PowHanSha11}.

The directions of the EM torque at these two resonances are also opposite. For the symmetric mode, $\theta = 0^{\circ}$ corresponds to the configuration of highest potential energy (unstable point), and thus the two repel each other once $\theta > 0 ^{\circ}$, until they come to the stable state at $\theta = 180 ^{\circ}$, while the reverse is true for the antisymmetric mode \cite{povinelli2005evanescent}. The evaluated external torque is about one order of magnitude smaller than the internal torque, and the total torque is of the order of $10^{-10}$\,Nm when the structure is pumped with a power density $P_\mathrm{I}$=1\,mW/mm$^{2}$. This is confirmed by the full-wave simulation (CST Microwave Studio) followed by calculation based on the Maxwell stress tensor, which yields the EM torque through a surface integral of the field components around the object \cite{jackson1999classical}.

The overall mechanism of achieving a nonlinear effect is presented in Figs.~\ref{fig:bist}(c,\,d). As an example, we choose a pump frequency (3.5\,GHz) at the symmetric mode [regime denoted by the black dashed line in Fig.~\ref{fig:bist} (b)]. It can be seen that $\mathbf{M}_\mathrm{EM}$ is a Lorentz-like function of the twist angle, while the restoring torques $\mathbf{M}_\mathrm{R}$ under different initial twist angles $\theta_0$ are approximated by linear functions (Hooke's law). The intersections of these two functions, $\mathbf{M}_\mathrm{EM}(\theta_e, P_\mathrm{I})+\mathbf{M}_\mathrm{R}(\theta_e) = 0$, correspond to the equilibrium angles $\theta_e$. However, only the angles with $\frac{\partial}{\partial \theta}\left[\mathbf{M}_\mathrm{EM}(\theta_e)+ \mathbf{M}_\mathrm{R}(\theta_e)\right] < 0 $ are stable. As the pump power $P_\mathrm{I}$ increases from zero to maximum and then reduces, the stable angles also change accordingly. With this method, we can numerically find a sequence of stable angles 
under different pump power $P_
\mathrm{I}$. 

Since the EM torque is a nonlinear function of twist angle, it naturally leads to nonlinear solutions. As shown in Fig.~\ref{fig:bist}(d), the power-dependent twist angles under different initial angles demonstrate the evolution from smooth nonlinear to bistable response as $\theta_0$ departs from the angle of maximum EM torque. In principle, as $\theta_0$ moves further away from the resonance, more noticeable rotation and hysteresis effects are expected, but higher pump power is required (see the case for $\theta_{0}=45^{\circ}$). Such evolution of the power-dependent nonlinear response can also be observed by fixing the initial twist angle but changing the pump frequency, as will be demonstrated in the experiment below. 

\begin{figure}[t]
\includegraphics[width=0.99\columnwidth]{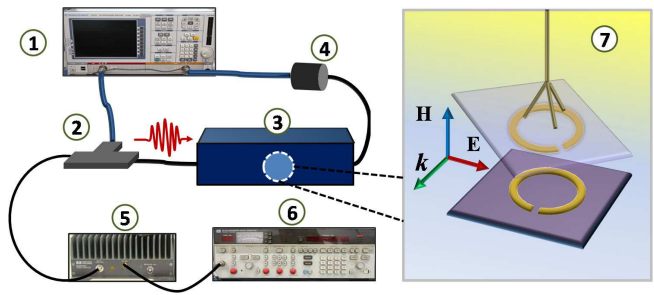}
\caption{Schematic of the experimental set-up.
The pump and probe signals are combined by a 3\,dB combiner.
The sample is positioned in the centre of the waveguide.
1: vector network analyser; 2: 3\,dB combiner; 3: rectangular waveguide;
4: 20\,dB attenuator; 5: power amplifier; 6: signal generator; 7: sample.
\label{fig:setup} }
\end{figure}

%--------------------------------------------------------
% \subsection*{Experimental verification}
%--------------------------------------------------------

To confirm the feasibility of the proposed nonlinear rotatable meta-atoms we carry out {\em a pump-probe microwave experiment}. To experimentally realise a strong nonlinear or even bistable effect, the restoring torque from the wire has to be sufficiently small so that the structure can be twisted by a large enough angle within the maximum available power. We found that rubber is a good candidate, since the shear modulus of rubber is of the order of 0.2\,MPa -- 2.4\,MPa \cite{gent2001engineering}, which is at least three orders of magnitude smaller than it is for other polymers. 

A schematic of the experimental setup is shown in Fig.~\ref{fig:setup}.
We have used two separated copper SRRs (inner radius $r=3.2$\,mm, track width 1\,mm, copper thickness $0.035\,\mu$m and slit width $g=0.2$\,mm) printed on Rogers R4003 substrates ($\epsilon_{\mathrm r}=3.5$, loss tangent 0.0027, substrate thickness 0.5\,mm). The lower SRR is fixed and positioned at the centre of a WR229 rectangular waveguide with $\Phi = 0^{\circ}$, and the top SRR is suspended with a thin rubber wire (radius $a=50\mu$m, length $d=20$\,mm), so that it can rotate about the common axis. The two SRRs are aligned coaxially, with a face to face distance of 0.75\,mm, separated by air. The horizontal positions of the SRRs are carefully adjusted and the initial twist angle $\theta_0$ is set at around 70$^{\circ}$. The mass of the suspended sample is 101\,mg, which leads to a $6.1\%$ elongation of the wire.
The Young's modulus is thus estimated as $2.06$\,MPa, and the shear modulus follows as $G \approx 0.69$\,MPa. The experimental realization varies slightly from Fig.~\ref{fig:scheme}, however the underlying physical mechanism of the nonlinear response due to the dynamic coupling between electromagnetic and elastic properties is the same. This equivalence is verified in the supplemental material \cite{supp}.

\begin{figure}[t]
\includegraphics[width=0.99\columnwidth]{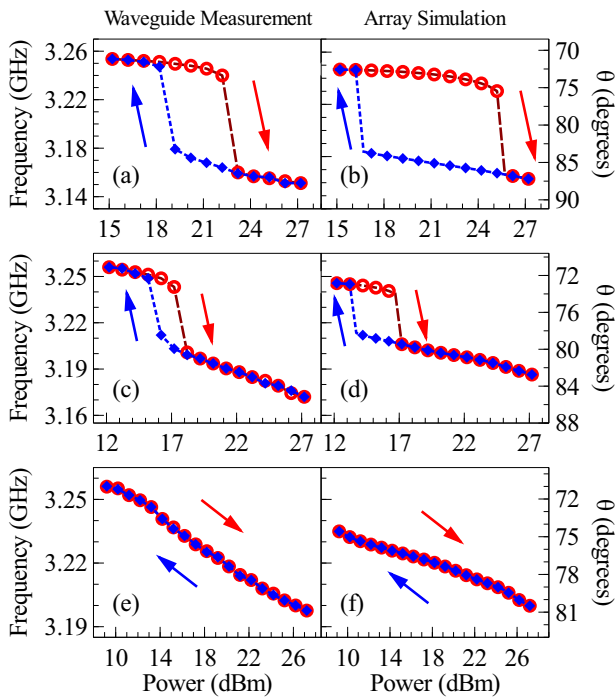}
\caption{Comparisons of resonant frequencies demonstrated in a waveguide experiment and numerically calculated for an array. (a) and (b) pump at 3.18\,GHz; (c) and (d) pump at 3.21\,GHz; (e) and (f) pump at 3.23\,GHz. The stable twist angles for the array system are shown on the right axes. The power in the array is the power incident on each unit cell.
\label{fig:comparison} }
\end{figure}

The transmission spectrum measurements are performed by a vector network analyser (Rohde and Schwarz ZVB-20). The CW pump signal is generated by a signal generator (HP 8673B) and is further amplified by a power amplifier (HP 83020A) before being sent into the waveguide. Three pump frequencies (3.18, 3.21 and 3.23\,GHz) are chosen in order to capture the evolution of the nonlinear response at different distances from the initial resonance. The pump power is increased in 1\,dB steps; for each step, the sample reaches steady state after 30 seconds. The mechanical rotation is quite significant and can be visually observed in the experiment, thus ruling out other possible nonlinear mechanisms such as heating. 

The experimentally observed transmission spectra are shown in Fig. S2 of the supplementary file, and the corresponding resonant frequencies are depicted in Fig.~\ref{fig:comparison}(a), (c) and (e), where the predicted evolution from bistability to smooth nonlinearity is clearly shown. The initial resonance (symmetric mode) without pump is located around 3.256\,GHz, and it red-shifts as the pump power increases, which indicates that the twist angle is increased. When the pump frequency is at the red tail of the resonance, a large spectral ``jump'' (about three times of the resonance linewidth) can be observed when the pump power passes a certain threshold value [Fig.~\ref{fig:comparison}(a)]. The thresholds are different for increasing and decreasing pump powers. As the pump frequency approaches to the initial resonance, the spectral ``jump'' becomes smaller [Fig.~\ref{fig:comparison}(c)] and finally disappears [Fig.~\ref{fig:comparison}(e)]. We also observed similar effects (not shown) when the pump 
frequency is at the red tail of the antisymmetric mode, in which case the two resonances approach each other due to the opposite direction of the EM torque. 

To further validate the observed effect, we numerically simulated the exact experimental geometry and found excellent agreement\cite{supp}. The estimated optimum initial twist angle is around $72.5^\circ$, and the maximum twist angles obtained for the three pump frequencies (3.18\,GHz, 3.21\,GHz and 3.23\,GHz) are around $90^\circ$, $85^\circ$ and $82^\circ$, respectively \cite{supp}. Finally we remap these angles back to the corresponding resonant frequencies according to the simulation spectra (see Fig. S3 (b), (d) and (f) of supplementary file).

%-------------------------------------------------------------
% \section*{Discussion}
%-------------------------------------------------------------
The single meta-atom used in the waveguide experiment induces image currents in the waveguide walls. Thus, our experimental system is analogous to an array where neighbouring elements interact. Since the nonlinearity arises from individual rotatable meta-atoms, the behaviour in an array will not show qualitative difference from a single meta-atom as predicted in the analytical model, as long as the neighbour interaction is relatively weak. We use full numerical simulation to model an array with thickness of a single cell and periodicity of 45\,mm in the transverse directions, arranged as in Fig.~\ref{fig:scheme}. The parameters of the meta-atoms used in the simulation are the same as in the experiment. The array and the waveguide experiment show quite good qualitative agreement (see Fig.~\ref{fig:comparison}(b), (d) and (f) for the power-dependent resonance and Fig. S1 (b) for the EM torque), thus 
justifying that the nonlinear behaviour observed in the waveguide experiment is similar to that of the analogous array system.

The novel nonlinear ``meta-atom'' described in this work proved to possess a sensitive elastic
feedback bringing nonlinearity to the interaction of EM modes of the resonators.
The resulting nonlinearity and bistability of the response was successfully observed in experiments
and it turns out that these results can be accurately predicted with theoretical modelling. We also note that this structure is chiral (except in the high symmetry cases of $0^\circ$ and $180^\circ$ angle between the rings), it should also exhibit nonlinear optical activity, an effect which is relatively weak in natural media \cite{kielich1967natural}, but can be quite strong in metamaterials \cite{shadrivov2011electromagnetic}. 

Although the experimental demonstration in this work was performed in the microwave frequency range,
the general principle of operation is valid at any frequency where a resonant response can be excited
in such or similar metamaterials elements, and the way to analyse the same phenomena in THz or optical
range is conceptually the same.

We believe that this work provides a substantial contribution to the emerging area of optomechanical
and magnetoelastic metamaterials, and offers an efficient and convenient design for practical applications.

% \section*{Acknowledgements}

This work was supported by Australian Research Council.
The authors are grateful to A.\,A.~Sukhorukov for helpful discussions.

\end{document}